# A study of electronic transport through a nanoscale air film

Ramonika Sengupta
*Electrical Engineering Department*
*Pandit Deendayal Petroleum University*
Gandhinagar 382421 INDIA
ramonika.sengupta@gmail.com

Anand Anil
*Department of Physics*
*Indian Institute of Science Education and Research*
Bhopal 462066 INDIA
anandanil@iiserb.ac.in

Santanu Talukder
*Electrical Engineering & Computer Science Department*
*Indian Institute of Science Education and Research*
Bhopal 462066 INDIA
santanu@iiserb.ac.in

*Abstract*— This paper presents a simple model for predicting electrical conductivity of air with varying electrode separation and different moisture content present in air. Our system consists of a metallic thin film (Cu) coated sample and a needle tip in a box filled with air. A constant potential difference has been applied between the tip and the sample and electrical current is calculated for different air-gaps and for different humidity conditions using COMSOL Multiphysics software. The electric potential, electric field and current density in the region between the probe tip and sample have been presented. The numerical results showing the variation of terminal current as a function of gap separation and relative humidity are found to be analogous with theoretical prediction. This study will be useful for finding the optimal current in case of electric field induced scanning probe nanolithography techniques such as electrolithography [1].

*Keywords — Scanning probe lithography, Electrolithography, conductivity of air, humidity, COMSOL, tunneling current.*

## I. Introduction

The past few decades have seen the electronic devices becoming smaller and smaller. Nanotechnology has played an important role in reduction of the size of semiconductor devices. Lithography [1-6] has been routinely used to produce nanopatterns on semiconductors to fabricate these devices. Scanning Probe Lithography (SPL) is one of the most popular techniques for nanopatterning. SPL uses a microscopic or nanoscopic scanning probe to physically or chemically modify the structures in immediate proximity of the probe. 'Electrolithography' (ELG) is a novel scanning probe based patterning technique, reported recently [1]. In case of ELG technique three dimensional structures are manufactured using a negatively biased probe that is mechanically moved over a surface to produce microscale or nanoscale patterns. The main advantage of this technique is high resolution (<10 nm) nanopatterning. Furthermore, unlike most of the other lithography techniques, ELG has low cost and ease of operation since it can be performed under atmospheric conditions. However, at present, it cannot be used for large scale patterning due to the production with low throughput. Therefore, it is mainly used for fabrication of nanostructures on individual samples.

In case of ELG and many other electric field induced SPL techniques it is important to understand – how the air film, present in between the tip and sample, behaves while applying an electric field. Furthermore, we need to understand the role of humidity or moisture content of air in deciding air film conductivity. This paper presents a simple 3D model consisting of a pin probe and a thin film sample enclosed in an air box to be used for various SPL techniques including ELG. Various SPL techniques use different kinds of probe and sample structure. For example, in case of ELG, a conductive tip is used for patterning and sample needs to be coated with a thin Cr film layer. However, in the context of this paper, for simplicity we have taken the tip to be metallic (Cu), and sample is a Si substrate coated with pure Cu film. This way we neglected any unwanted effect arising from electronic properties of the tip or sample material. In addition, we could also avoid the oxidation effect on the tip or the sample. The system has been modelled using COMSOL Multiphysics version 5.4. The terminal current of the probe, which is crucial for precise nanopatterning, has been estimated for various probe tip diameters and for different separation gap between the probe and the sample. Moreover as the atmospheric conditions play an important role in determining the magnitude of probe terminal current, the effect of humidity on the terminal current has been also investigated.

## II. Pin probe Model for SPL in COMSOL

Fig. 1 shows a 3D schematic diagram of the model used in COMSOL platform for simulating electrical conduction through air. The system consists of a conical probe and a thin film sample. Both the probe and sample are kept in an air filled box. The principles of electrostatic theory have been used for simulating the electrical conduction through the scanning probe model shown in Fig. 1. Under the AC/DC module of COMSOL software, different constants, variables and materials are defined for the model. Similar to ELG technique, a negative potential is applied to the conical probe and the sample plate is kept at ground potential. Both of the

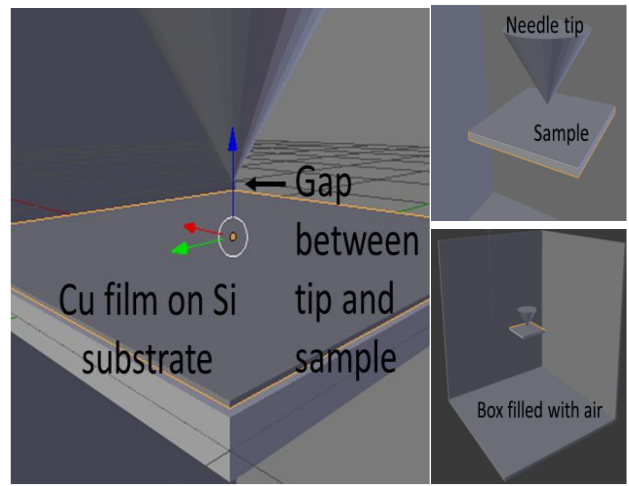

Fig 1: Schematic diagram of pin-probe and sample.

*Corresponding author: Santanu Talukder, e-mail: santanu@iiserb.ac.in*

conductors are enclosed within a hollow copper block to provide for the boundary conditions. To simulate atmospheric conditions the block enclosing the probe and plate is filled with air of specified conductivity.

### A. Geometry of pin probe model

The model consists of a probe and a plate. The probe has a bottom radius of 0.01 mm, top radius of 0.5 mm and a height of 1 mm. The probe is given –ve potential and is positioned in XY plane at z = 0.7 mm. The grounded plate, which serves as the sample for nanostructuring, is positioned in XY plane at z = 0 mm. The plate thickness has been taken as 0.2 mm. This entire set-up is enclosed inside a conductor box with dimensions of 10 mm x 10 mm x 10 mm to provide for the boundary conditions for the assembly.

### B. Materials used for set up

The material used for the probe, plate and the block is copper. The block used is a hollow copper block, containing air. The electrodes are surrounded by air and the effect of humidity is included by varying the conductivity of the air as a function of relative humidity [9], which in turn is related to the permittivity.

### C. Physics used in model

The electric current formulation of AC/DC module of COMSOL is used to estimate the potential distribution between two electrodes. The corresponding electric field in the gap is also determined. The electric current density is related to electric field through Ohm's law (i.e., $J = \sigma E$), hence the current density distribution is similar to the electric field distribution and is also determined by the software. Finally, the terminal current is obtained using the software by integrating the current density over the area. Extra fine Physics Controlled mesh was chosen for all the cases. Stationary and parametric analyses were performed.

### III. RESULTS AND DISCUSSION

Fig. 2 shows the plot of potential distribution in the region around pin probe. The corresponding 3D plot for electric field showing the field lines is given in Fig 3. The potential drop is very sharp in the vicinity the probe tip, hence the electric field should be very high around the tip. We have also obtained the electric potential distribution and electric field with varying probe tip diameters. As we decrease the diameter of the probe tip, the electric field becomes stronger. The electric potential also differs for different atmospheric conditions in the vicinity of the pin probe. At atmospheric pressure it has much sharper drop near the pin probe than that under vacuum. Consequently the electric field is sharper near the pin probe surrounded with air than under vacuum. The variation of current density is similar to the variation of electric field. This is as expected since the relationship $J = \sigma E$ should hold. Here, the conductivity of air has taken to be $10^{-9}$ S/m for these plots.

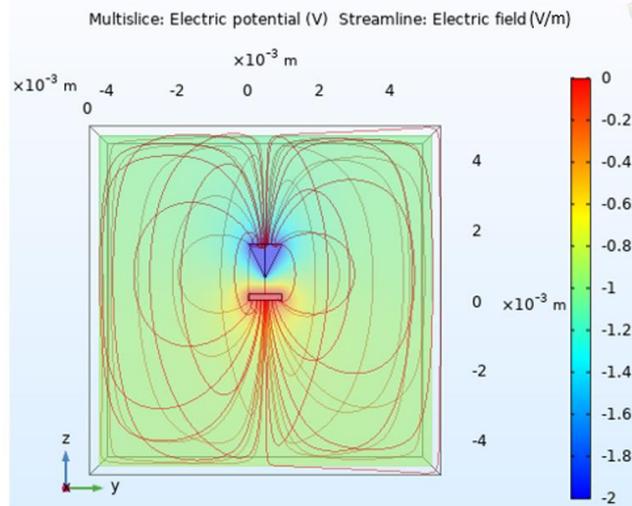

Fig. 3: Streamline plot showing electric field lines in the region around probe

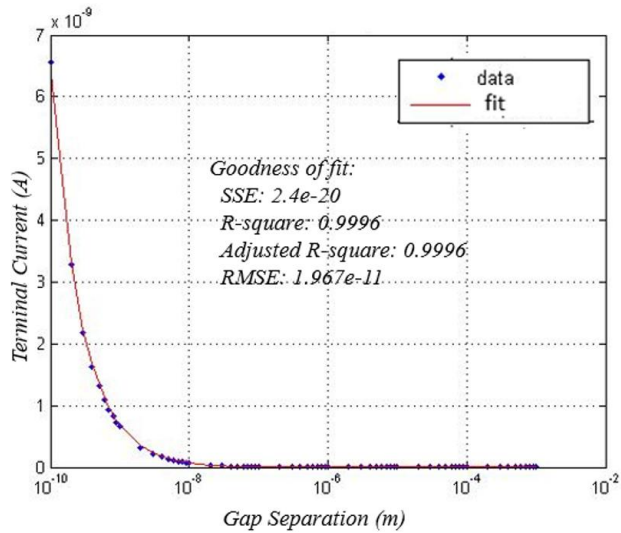

Fig 4: Terminal current vs gap separation (*0.1nm <z< 1mm*) for probe tip diameter *d = 20 μm*, the dots show COMSOL simulation and continuous line is empirical fit.

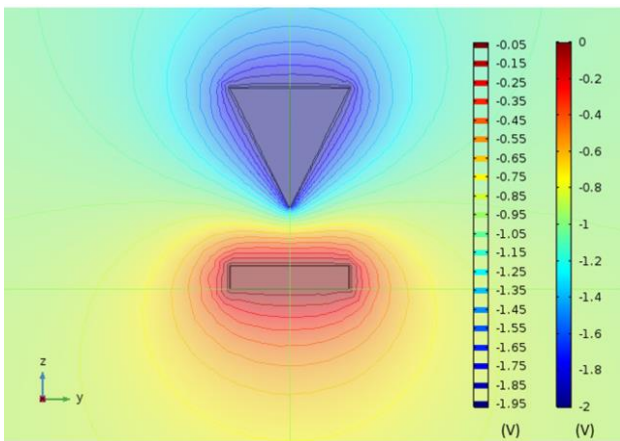

Fig 2: Potential distribution (in V) in the region around the probe

The terminal current or total current passing between the pin probe and the grounded sample is then obtained by integrating the current density over the collection area. The gap of separation between the pin probe and the grounded plate was varied and the corresponding terminal current was obtained. Fig. 4 shows the terminal current as a function of gap separation *z* in the range of (*0.1nm <z< 1mm*) for probe diameter of *d = 20 μm*. An empirical formula is obtained for the terminal current *I(z)* as a function of the gap separation *z* for probe diameter of *d = 20 μm* and compared with the data obtained from COMSOL to obtain the coefficients. Current should exponentially decay with increasing the gap. Because, in this nanoscale gap it follows tunnelling phenomenon. So, we can fit the result according to the equation

$$I(z) = a\, e^{bz} \qquad (1).$$

Here, $a$ and $b$ are the constants having values $1.76\times10^{-12}$ and $-4.62$ respectively and $z$ is normalized by the mean. Both the COMSOL simulation and the empirical fit are shown in Fig. 4. As shown in Fig. 4, simulation results and theoretical prediction have very good match ($R_{sq} = 0.996$) with each other.

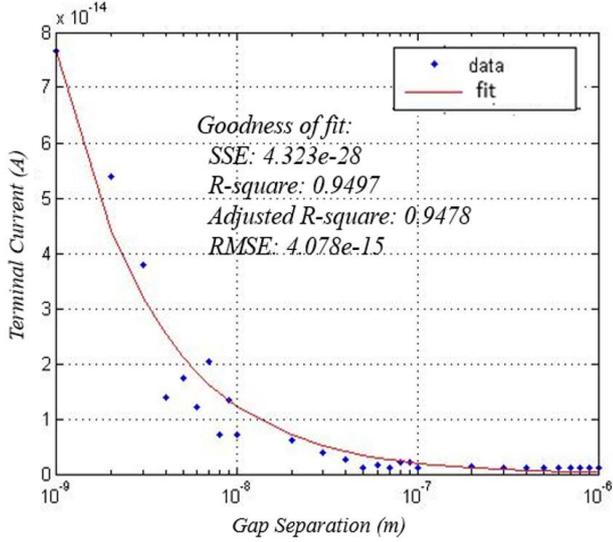

Fig 5: Terminal current as a function gap separation with probe tip diameter of 10 nm, the dots show COMSOL simulation and continuous line is empirical fit.

Fig. 5 shows the similar plot of terminal current as a function of gap separation for much smaller probe diameters i.e., $d = 10\ nm$. The empirical fit to the data is also shown in the figure. Here the values of the constants $a$ and $b$ are $5.14\times10^{-21}$ and $-0.80$ respectively.

We have further studied the effect of atmospheric conditions on the terminal current. Electrolithography is performed under atmospheric conditions, i.e., the temperature and humidity of the air between the electrodes may vary. Precise knowledge of the terminal current is required for nano- or micro-patterning under these varying ambient conditions. However, since the conductivity of air changes with the relative humidity [9], as a result, the terminal current also changes correspondingly. Hence we have estimated the variation of terminal current with conductivity which translate to the variation of terminal current with relative humidity of the air. Fig. 6 shows the behaviour of terminal current as a function of saturation ratio s which is the measure of relative humidity of air (i.e., s = %RH/100). The fitted empirical relation is also shown in the figure. The behaviour of the terminal current with saturation ratio s is given by the following empirical formula

$$I(s) = c\, s^{13} \qquad (2)$$

where c is the constant empirically determined from the terminal current curve as $1.62\times10^{-15}$. This simulation result also matches with the theoretical prediction made in Ref. [9].

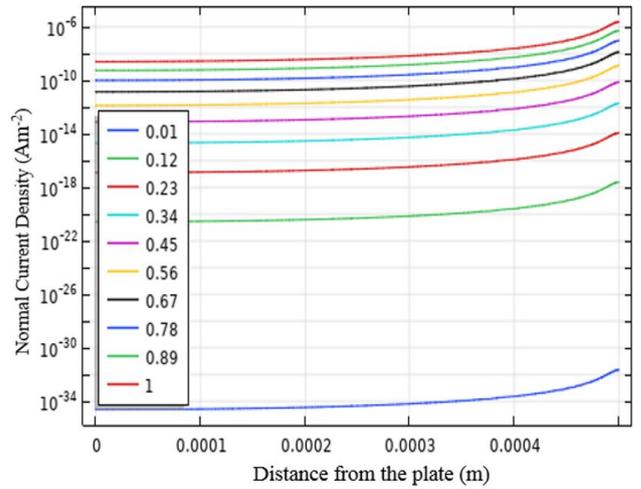

Fig 7: Current Density as function of gap distance, along line joining grounded plate and probe tip for different values of saturation ratio s

Fig. 7 shows the variation of current density with gap separation for various values of saturation ratio s. The current density increases at all values or gap separation with increasing humidity in air.

## IV. CONCLUSIONS AND SUMMARY

The modelling of pin probe system for ELG technique has been performed in this work. The current flowing between the probe tip and the substrate is very crucial for precise nano-patterning on a substrate. We have estimated the terminal current for different probe tip diameters and with varying gap separations between the tip and sample. Current through the air film has been seen to decrease exponentially with increasing gap separation between the tip and the sample. Furthermore, since the ELG technique is done at atmospheric conditions, the humidity plays an important role in determining the terminal current. Hence we have estimated the effect of relative humidity of air on the magnitude of current density. It has been seen that the current density, hence the terminal current increases with increasing relative humidity. This study will be helpful to optimize the current for different scanning probe based patterning techniques, which are performed in ambient conditions.

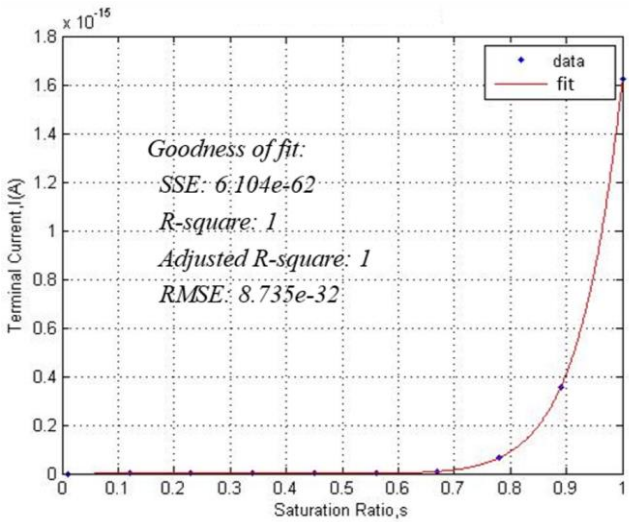

Fig. 6: Terminal current as a function of saturation ratio s

ACKNOWLEDGMENT

Authors wish to thank MHRD, Govt. of India for financial support, and Mr. Aryan Chaudhary and Mr. Anirban Sardar for fruitful discussions.